\documentclass{article}

    \PassOptionsToPackage{numbers, compress}{natbib}
\usepackage{tackling_climate_workshop_style}

\usepackage[utf8]{inputenc} % allow utf-8 input
\usepackage[T1]{fontenc}    % use 8-bit T1 fonts
\usepackage{hyperref}       % hyperlinks
\usepackage{url}            % simple URL typesetting
\usepackage{booktabs}       % professional-quality tables
\usepackage{amsfonts}       % blackboard math symbols
\usepackage{nicefrac}       % compact symbols for 1/2, etc.
\usepackage{microtype}      % microtypography
\usepackage{amsmath} 
\usepackage{graphicx}
\usepackage[toc,title,page]{appendix}
\usepackage{array}
\usepackage{float}

\title{Inference of CO$_2$ flow patterns--a feasibility study}

\author{%
  David S.~Hippocampus\thanks{Use footnote for providing further information
    about author (webpage, alternative address)---\emph{not} for acknowledging
    funding agencies.} \\
  Department of Computer Science\\
  Cranberry-Lemon University\\
  Pittsburgh, PA 15213 \\
  \texttt{hippo@cs.cranberry-lemon.edu} \\
}

\begin{document}

\maketitle

\begin{abstract}

As the global deployment of carbon capture and sequestration (CCS) technology intensifies in the fight against climate change, it becomes increasingly imperative to establish robust monitoring and detection mechanisms for potential underground CO${_2}$ leakage, particularly through pre-existing or induced faults in the storage reservoir's seals. While techniques such as history matching and time-lapse seismic monitoring of CO${_2}$ storage have been used successfully in tracking the evolution of CO${_2}$ plumes in the subsurface, these methods lack principled approaches to characterize uncertainties related to the CO$_2$ plumes' behavior. Inclusion of systematic assessment of uncertainties is essential for risk mitigation for the following reasons: (i) CO$_2$ plume-induced changes are small and seismic data is noisy; (ii) changes between regular and irregular (e.g., caused by leakage) flow patterns are small; and (iii) the reservoir properties that control the flow are strongly heterogeneous and typically only available as distributions. To arrive at a formulation capable of inferring flow patterns for regular and irregular flow from well and seismic data, the performance of conditional normalizing flow will be analyzed on a series of carefully designed numerical experiments. While the inferences presented are preliminary in the context of an early CO$_2$ leakage detection system, the results do indicate that inferences with conditional normalizing flows can produce high-fidelity estimates for CO$_2$ plumes with or without leakage. We are also confident that the inferred uncertainty is reasonable because it correlates well with the observed errors. This uncertainty stems from noise in the seismic data and from the lack of precise knowledge of the reservoir's fluid flow properties.
%
% Consequently, it becomes necessary to incorporate observations derived from the geophysical time-lapse observations into these simulations and account for their uncertainties which can come from the data noise, initial saturation state, and the permeability model. In this study, we introduce an approach that harnesses conditional normalizing flow for the prediction of CO${_2}$ plumes, with a focus on conditioning these predictions on geophysical time-lapse observations. 
% This methodology represents a pivotal step towards enhancing our ability to predict and monitor CO${_2}$ leakage in the context of CCS technology, thereby ensuring its safety and long-term sustainability.
%
\end{abstract}
\section{Introduction}
According to the International Panel on Climate Change 2018 report \cite{IPCC2018}, achieving a 50$\%$ reduction in greenhouse gas emissions by the year 2050 to avert a 1.5-degree Celsius increase in the Earth's average temperature is critical. It entails large-scale deployment of carbon reduction technologies, most notably carbon capture and storage (CCS). CCS involves the collection, transportation, and injection of carbon dioxide (CO${_2}$) into suitable underground geological storage sites. This long-term storage process, known as Geological Carbon Storage (GCS), ranks amongst the scalable net-negative CO$_2$ emission technologies. However, the viability of GCS is contingent on mitigating risks of potential CO${_2}$ leakage from underground reservoirs, which can result from pre-existing fractures in reservoir seals, as underscored in the study by \cite{ringrose2020store}. For this reason, there is a pressing need to mitigate these risks by instituting robust monitoring systems capable of accurate prediction of subsurface CO${_2}$ plume behavior.

Recently, several methods have emerged that leverage machine learning to detect CO${_2}$ leakage within CO$_2$ storage complexes \cite{LI2018276, ZHOU2019102790, erdinc2022AAAIdcc, yin2022TLEdgc}. However, these techniques do not provide information on the spatial extent of leakage and its uncertainty. Despite these shortcomings, advanced generative models have been deployed successfully to predict the dynamic evolution of CO${_2}$ plumes based on saturation and pressure data collected at the well(s) \cite{https://doi.org/10.1029/2018WR024592, STEPIEN2023204982}.  In this work, we conduct a preliminary study to demonstrate how observed multi-modal (well and seismic) time-lapse monitoring data can be used to improve inferences of both regular and irregular (e.g., due to leakage) CO$_2$-flow patterns including quantification of their uncertainty. To carry out these inferences, we referred \cite{siahkoohi2022ravi} which showed that Conditional Normalizing Flows (CNFs) can approximate posteriors of seismic imaging. We also make use of CNF's ability to handle non-uniqueness \cite{siahkoohi2022ravi}, an essential capability when dealing with a nonlinear phenomena. While the presented reservoir and seismic simulations are realistic, this paper can only be considered as an early attempt to demonstrate CNF's ability to capture subtle differences between regular and irregular flow patterns and their uncertainty from multi-modal time-lapse data.
\section{Methodology}
\begin{figure*}[t]
\centering
\includegraphics[width=0.80\textwidth] {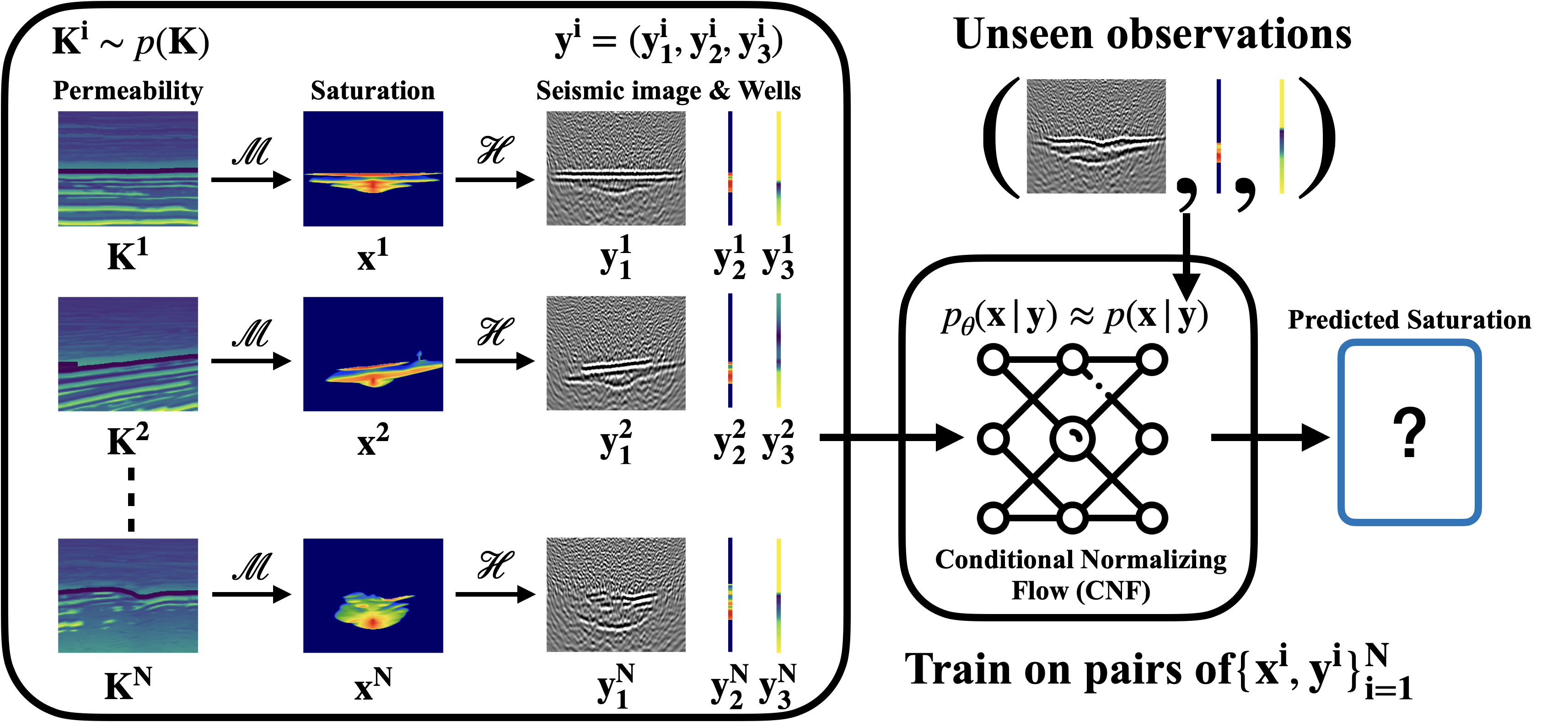}
\caption{\small{Creation of training and testing data for the conditional normalizing flow. Training pairs are simulated by running fluid flow simulations ($\mathcal{M}$) for random samples of the permeability ($\mathbf{K}^{(i)}\sim p(\mathbf{K}),\, i=1\cdots N$). These reservoir simulations produce samples for the CO$_2$ saturation, $\mathbf{x}^{(i)},\, i=1\cdots N$, and pressure. The simulated plumes are observed, via the observation operator $\mathcal{H}$, directly at the well, producing localized measurements for the saturation and pressure, and indirectly via imaged seismic reflection data collected at the surface. The conditional normalizing flow is trained on pairs of CO$_2$ saturation and corresponding time-lapse observations collected in the vector $\mathbf{y}^{(i)},\, i=1\cdots N$. Each $\mathbf{y}^{(i)}=(\mathbf{y}^{(i)}_1,\mathbf{y}^{(i)}_2,\mathbf{y}^{(i)}_3)$ with the subscripts denoting seismic (1), saturation at well (2), and pressure at well (3), respectively. After training, the normalizing flow is tested on unseen observations.}}\label{fig:figure1}
\end{figure*}
\subsection{Conditional Normalizing Flows}
Normalizing flows are generative models that approximate complex target distributions by applying a series of invertible and differentiable transformations ($f_{\theta}: \mathbb{R}^d \rightarrow \mathbb{R}^d$ \text{ with inverse} $f_{\theta}^{-1}$) on a base known distribution (Normal distribution)\cite{nf}. After training, normalizing flows can generate samples from the target distribution by performing the inverse operation on the base distribution. Since we want samples from the conditional distribution, we utilize conditional normalizing flows \cite{cnf} where the mapping from the base density to the output space is conditioned on time-lapse observations $\mathbf{y}$ to model the posterior distribution of CO$_2$ saturation images, denoted as $p\left(\mathbf{x}\mid\mathbf{y}\right)$, with $\mathbf{x}$ being CO$_2$ saturation image and $\mathbf{y}$ being time-lapse observables. The training objective is
\begin{equation}\label{eq1}
\resizebox{0.93\hsize}{!}{%
    $\mathcal{L}({\theta}) = \mathbb{E}_{\mathbf{x},\mathbf{y} \sim p(\mathbf{x}, \mathbf{y}) } \left[ \frac{|| f_{\theta}(\mathbf{x};\mathbf{y})||_{2}^{2} }{2} - \log|\det(\mathbf{J}_{f_{\theta}}(\mathbf{x}, \mathbf{y}))|  
    \right] 
    \approx 
    \frac{1}{N} \sum^{N}_{i=1} \left( \frac{|| f_{\theta}(\mathbf{x^{(i)}};\mathbf{y^{(i)}})||_{2}^{2} }{2} - \log|\det(\mathbf{J}_{f_{\theta}}(\mathbf{x}^{(i)}, \mathbf{y}^{(i)}))| \right),$%
    }
\end{equation}
where $\mathbf{J}$ represents the Jacobian of the network $f_{\theta}$ with respect to its input. This training objective corresponds to minimizing the Kullback-Leibler divergence between the target density and the pull-back of the standard Gaussian distribution defined by the invertible network\cite{siahkoohi2023reliable,orozco2023MIDLanf}. The expectation is approximated by an average of $N$ training samples. After training, posterior samples of saturation images are generated by applying the inverse transformations to random Normal noise realizations conditioned on the observed geophysical data. These posterior samples serve as a basis for statistical analyses, including estimating the posterior variances to assess the uncertainty and to make a high-quality point estimate. We use the posterior mean calculated by the routine in Appendix \ref{training}.
\subsection{Dataset Generation}
We select 850 2D vertical slices derived from the 3D Compass velocity model \cite{BG}, to create the training dataset for our conditional normalizing flows. This model, though synthetic, is obtained from real seismic and well-log data and, thus, emulates the realistic geological characteristics prevalent in the southeastern region of the North Sea. Each 2D slice corresponds in physical dimensions to $2131\mathrm{m}\times 3200\mathrm{m}$. To simulate the dynamics of CO${_2}$ flow, we follow \cite{yin2023solving} and convert the velocities of the Compass model \cite{BG} to models of the permeability and porosity using empirical relationships including the Kozeny-Carman equation \cite{https://doi.org/10.1029/2005GL025134}. The flow simulations are carried out with the open-source packages Jutul.jl \cite{jutul} and JutulDarcyRules.jl \cite{jutuldarcyrules} while seismic data modeling and imaging are done with JUDI \cite{judi}, which is a Julia front-end to Devito \cite{gmd-12-1165-2019, 10.1145/3374916}, a just-in-time compiler for industry-scale time-domain finite-difference calculations. Next, fluid flow and wave simulations are briefly discussed. Refer to \cite{louboutin2023lmi} for more detail on the numerical simulations.
\subsubsection{Fluid Flow Simulations}
To obtain a realistic CO$_2$ injection, an injectivity of 1 MT/year is chosen with vertical injection intervals inside the high permeability regions. As CO${_2}$ is injected supercritically, the CO$_2$ saturations and pressures are calculated by numerically solving the equations for two-phase flow. Details on these numerical solutions of the partial-differential equations can be found in \cite{RASMUSSEN2021159}. Two distinct flow scenarios, namely regular flow (no-leakage) and irregular flow (leakage), are considered. During no-leakage, the reservoir properties are  kept constant resulting in regular CO${_2}$ plumes. However, leakage occurs when the permeability changes at the reservoir's seal, which results in an irregular flow. While leakage can be caused by many mechanisms, we only consider the one due to pressure-induced opening of pre-existing fractures in the seal. In this case, leakage is triggered when CO${_2}$ injection pressure reaches a predefined threshold \cite{ringrose2020store}, resulting in an instantaneous permeability increase within the seal causing the CO${_2}$ plume to leak out of the reservoir. To train the CNFs, 1700 multiphase flow simulations are performed ($N=1700)$, 850 with and 850 without leakage. In practice, these fluid flow simulations can also be performed by computationally cheap surrogates based on model-parallel Fourier neural operators \cite{grady2023model}, enhancing its adaptability to large-scale four-dimensional scenarios. In the next section, we describe the formation of seismic images of these regular/irregular plumes.

\subsubsection{Time-lapse Seismic Imaging}
As injection of CO${_2}$ induces changes in the Earth's acoustic properties (velocity and density), these changes can be observed seismically. To mimic the process of collecting time-lapse seismic data followed by imaging, seismic baseline and monitor surveys are modeled. During these simulations, the baseline survey represents the initial stage before CO${_2}$ is injected, and the monitor survey corresponds to the time of 8 years after the injection. The seismic acquisition uses 8 sources and 200 ocean bottom nodes, along with a 15 Hz Ricker wavelet and a band-limited noise term with a signal-to-noise ratio (SNR) of 8.0 dB. Reverse time migration (RTM) \cite{doi:10.1190/1.1441434} is employed to create time-lapse seismic images of the subsurface. Then, we isolate the time-lapse changes attributed to CO${_2}$ saturation by subtracting the baseline and monitor images.
\section{Training and Results}
To create training pairs, $\{ \mathbf{x}^{(i)}, \mathbf{y}^{(i)} \}_{i=1}^{1750}$, we resize the saturation dataset $(\mathbf{x})$ into a 256 $\times$ 256 single channel images and the time-lapse data $(\mathbf{y})$ into 256 $\times$ 256 $\times$ three channel images. The three channels are the imaged seismic observations, pressure well, and saturation well data, respectively. The architecture of the conditional normalizing flow is similar to \cite{cnf}. Refer to Appendix \ref{training} for further details and hyperparameter selection.

\begin{figure*}[t]
\centering
\includegraphics[width=0.8\textwidth]{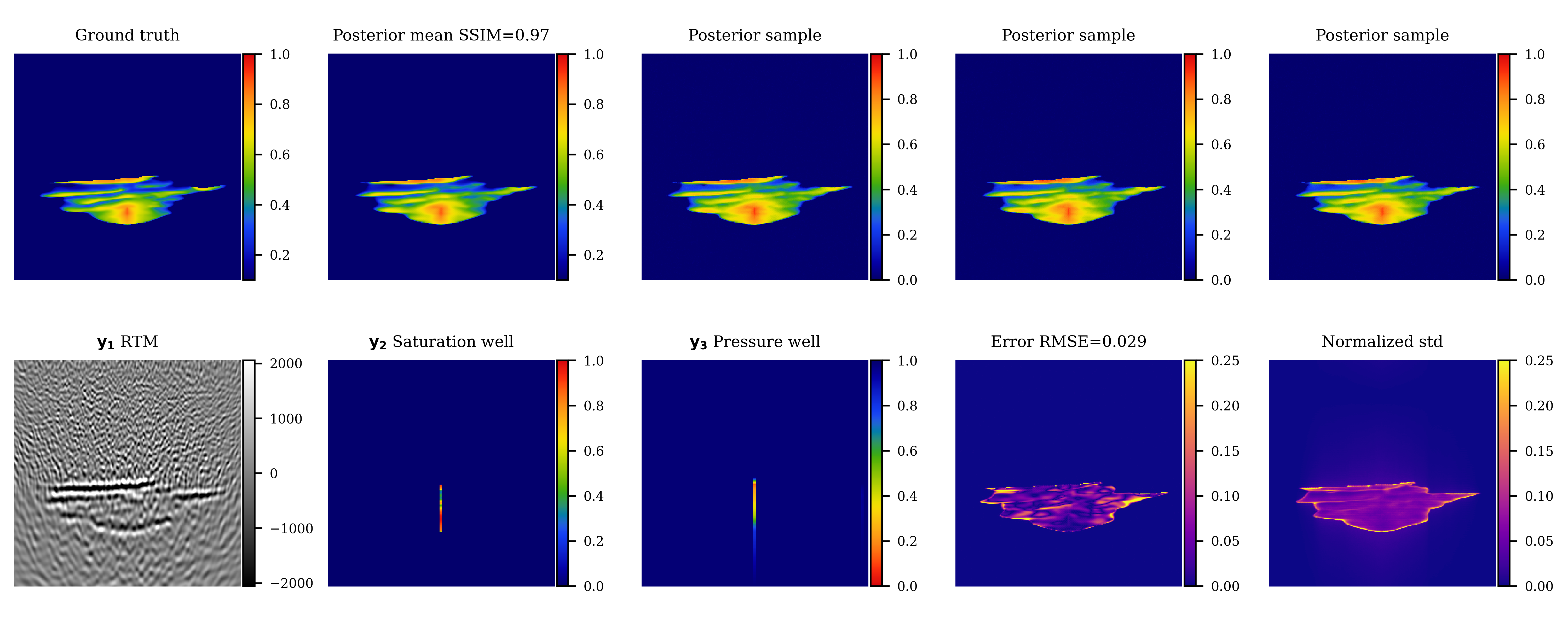}
\caption{\small{Outputs from the trained network for no-leakage case. Refer to Appendix \ref{examples} for details on performance metrics and additional examples.}}
\label{fig:figure2}
\end{figure*}

\begin{figure*}[t]
\centering
\includegraphics[width=0.8\textwidth]{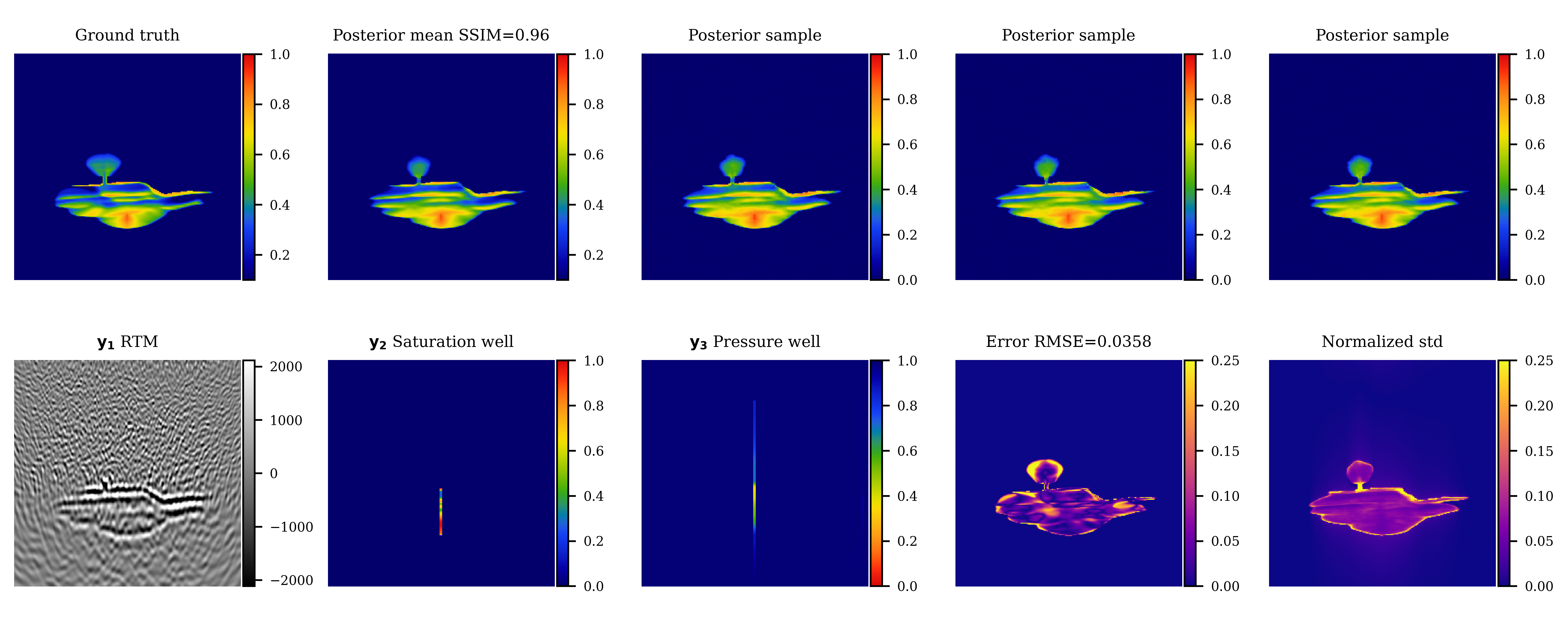}
\caption{\small{Same as Figure \ref{fig:figure2} but now for leakage case.}}
\label{fig:figure3}
\end{figure*}

After training, the conditional normalizing flow generates samples from the posterior distribution of CO$_{2}$ saturation given unseen seismic and well observations. Figure \ref{fig:figure2} \& Figure \ref{fig:figure3} show the outputs for a no-leakage case and leakage case, respectively. The posterior means of the samples appear close to the ground truth and have SSIM (see Appendix \ref{examples}) values of $0.97$ and $0.96$ for the no-leakage and leakage case. As expected, the uncertainty (normalized std) is higher in geologically complex areas such as the top of the plume, which corresponds to the bottom of the seal and the fracture region from where CO$_{2}$ leaks out and it also correlates well with the error. We show more test samples in Appendix \ref{examples}.  Although there are errors in our method's capability to find the exact extent of the plume, we do not observe any false positives or false negatives (positive and negative refer to leakage and no-leakage respectively) from the 36 test samples. In other words, all leakage scenarios are clearly inferred as leakage and all no-leakage scenarios are inferred as no-leakage. 
%
%% !!! Replace figure 2 !!!
\section{Conclusion and Discussion}
Monitoring of GCS is complicated by highly nonlinear relationships between the reservoir properties, the CO$_2$ plumes, and time-lapse seismic observations. These complications are compounded by the fact that the reservoir properties are only available statistically, making it difficult to detect potential CO$_2$ leakages that lead to subtly differing flow patterns. By employing carefully designed numerical experiments, we are able to demonstrate that conditional normalizing flows are capable of capturing these subtle pattern changes during inference in a setting where training pairs consist of realizations for the CO$_2$ saturation and associated time-lapse data, consisting of seismic images and well measurements, for scenarios that include regular and irregular (leakage) flow. Aside from producing estimates for the CO$_2$ saturation that only differ slightly from the ground truth, these inferences also produce estimates for the uncertainty that correlate well with the errors. In future work, we will study how this type of inference can lead to an uncertainty-aware ML-based monitoring system capable of early leakage detection. This feasibility study can also serve as an initial step towards constructing a digital twin of a geological carbon storage monitoring system that receives real-time data updates, and employs simulation, machine learning, and reasoning methodologies to facilitate decision-making processes. This can be achieved by employing sequential Bayesian inference of CO$_2$ plumes conditioned on time-lapse geophysical observations as discussed in \cite{herrmann2023dte}.

%Version2
% This study establishes the foundation for developing a digital twin for CCS, envisioned as a virtual representation of a carbon storage monitoring system equipped with real-time data updates. As we progress, our objective is to incorporate a sequential component into the project, drawing upon the concepts outlined in \cite{herrmann2023dte}.
%

\section{Acknowledgements}
This research was carried out with the support of Georgia Research Alliance and partners of the ML4Seismic Center. This research was also supported in part by the US National Science Foundation grant OAC 2203821.

\appendix
\renewcommand{\thesection}{\Alph{section}.\arabic{section}}
\setcounter{section}{0}

\bibliography{biblio.bib}

\begin{appendices}

\section{Training Setting} \label{training}\label{network_detail}
We use following hyperparameters during training experiment (see Table\ref{table:1}).

\begin{table}[th]
\centering
\begin{tabular}{ >{\centering\arraybackslash}m{1.5in}  >{\centering\arraybackslash}m{1.5in} }
\toprule[1.5pt]
\multicolumn{2}{c}{Training Hyperparameters} \\ \midrule[1.5pt]
Batch Size & 32 \\
Optimizer & Adam \cite{Kingma2014AdamAM} \\
Learning rate (LR) & $10^{-3}$ \\
No. of training epochs & 100 \\
Fixed Noise Magnitude & 0.005 \\
No. of training samples & 1632 \\
No. of validation samples & 68 \\
No. of testing samples & 36 \\
\bottomrule[1.5pt]
\end{tabular}
\caption{\small{Hyperparameter for the training experiment.}}
\label{table:1}
\end{table}

% To monitor the occurrence of overfitting, we assess the objective function at the end of each epoch over a random subset of the validation set. Figure \ref{fig:loss_graph}
% shows change in L$_{2}$ term, Logdet term, total objective (L$_{2}$ + Logdet), average SSIM value and average L$_{2}$ distance between ground truth and generated images after each parameter update respectively for both training and validation sets. 

% \begin{figure*}[h]
% \centering
% \includegraphics[width=\textwidth]{Appendix Figures/loss_graph.png}
% \caption{\small{The change in network objective and evaluation metric of SSIM and L$_{2}$ distance after each parameter update for train and validation}}
% \label{fig:loss_graph}
% \end{figure*}

After the completion of training, we use the following procedure to calculate posterior mean:
\begin{equation}
\resizebox{0.73\hsize}{!}{%
    $\mathbf{x}_{\text{PM}} = \mathbb{E}_{\mathbf{x} \sim p(\mathbf{x}|\mathbf{y})} [ \mathbf{x} ]
    \approx 
    \frac{1}{M}  \sum^{M}_{n=1}\mathbf{x}_{\text{gen}}^{i} \text{ where } \mathbf{x}_{\text{gen}}^{i} = f_{\hat{\theta}}^{-1}(\mathbf{z_i};\mathbf{y}) \text{ with } \mathbf{z_i} \sim \mathcal{N}(0,I),$%
    }
\end{equation} \label{posteriormean}
where $\hat{\theta}$ is the minimizer of Equation \ref{eq1}.

\section{Generated Examples and Useful Definitions} \label{examples}

% In figure \ref{fig:appendix_noleak} and \ref{fig:appendix_leak}, we show two more examples for scenarios of no-leakage and leakage respectively. 

\begin{figure*}[h]
\centering
\includegraphics[width=0.7\textwidth]{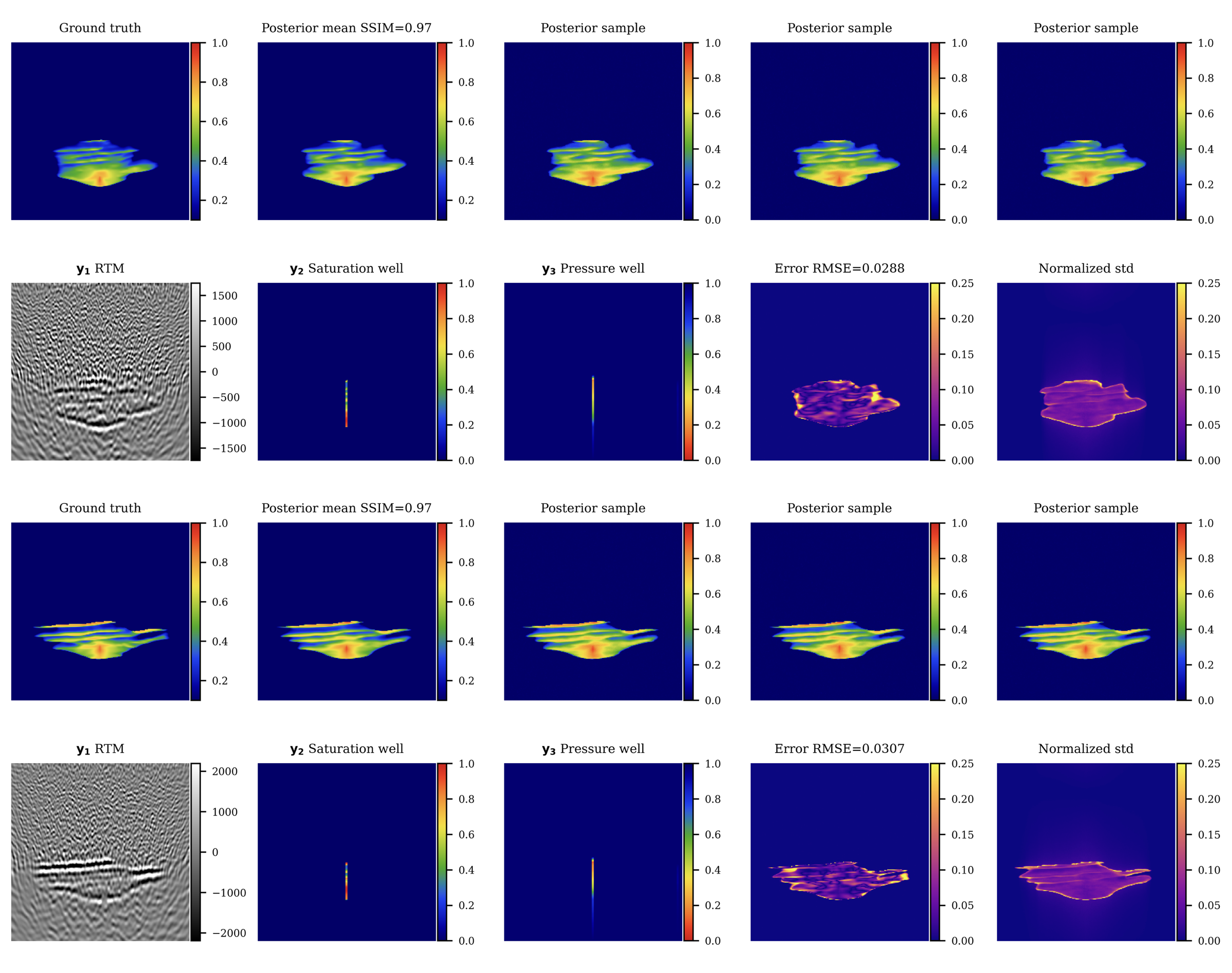}
\caption{\small{Two distinct examples for no-leakage scenario. In both cases, the network can generate high-fidelity saturation images with posterior mean SSIM of 0.97. Relatively larger uncertainty values are concentrated around the boundary of CO$_{2}$ plumes.}}
\label{fig:appendix_noleak}
\end{figure*}

\begin{figure*}[t]
\centering
\includegraphics[width=0.7\textwidth]{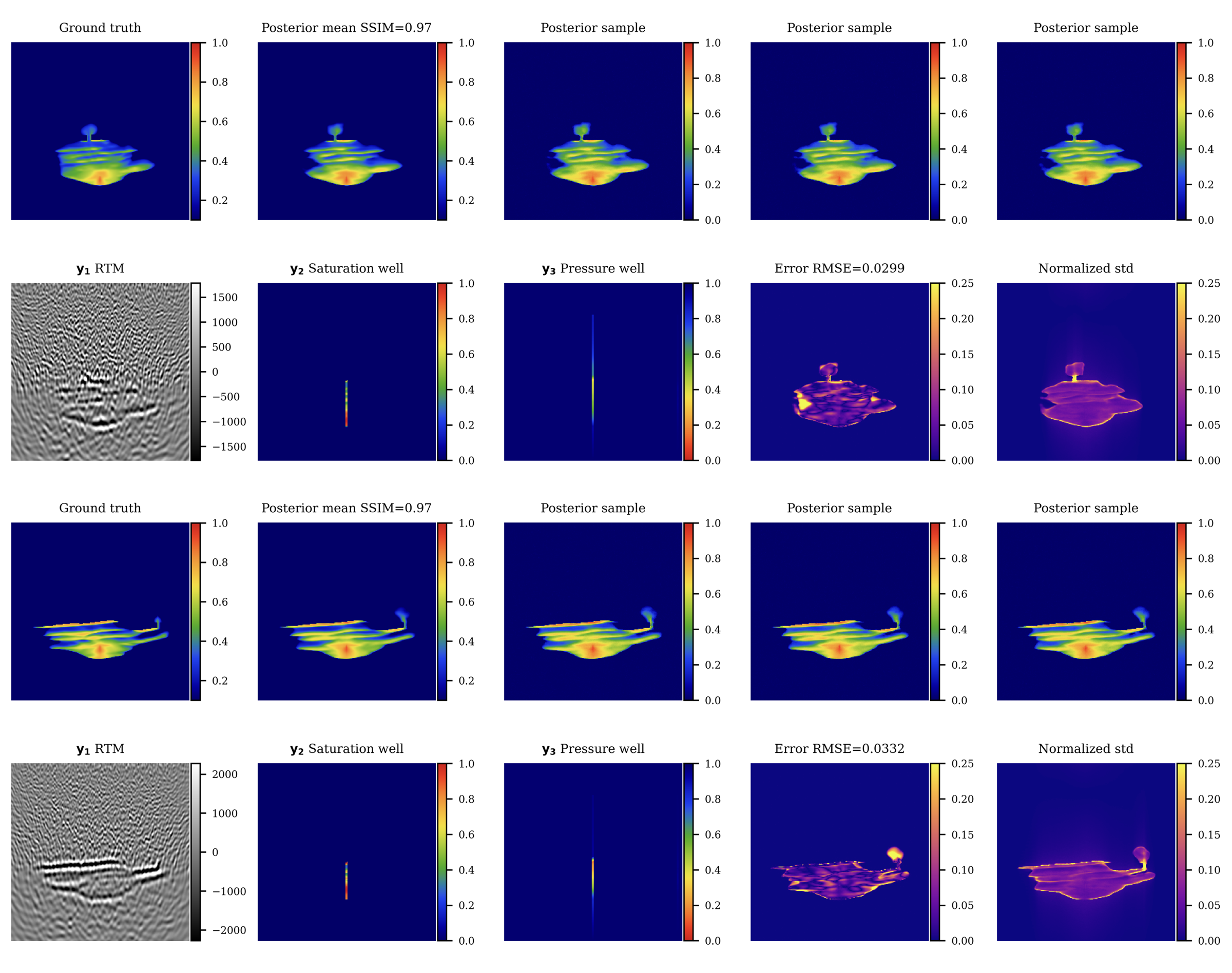}
\caption{\small{The corresponding leakage scenario for previous two examples. In both instances, the leakage posterior means have SSIM of 0.97. Relatively larger uncertainty values are concentrated within the area of the CO$_2$ leakage, particularly along the canal (fractured seal area) and the edges of the plume. }}
\label{fig:appendix_leak}
\end{figure*}

Metric definitions used in Figures \ref{fig:figure2}, \ref{fig:figure3}, \ref{fig:appendix_noleak} \& \ref{fig:appendix_leak} and text:

\textbf{SSIM} - Structural similarity index quantifies the similarity between two images and is commonly used to assess how closely a generated image resembles a ground truth or reference image. It considers image quality aspects such as luminance, contrast, and structure. For the mathematical formulation of SSIM, please refer to the study by \cite{1284395}.

\textbf{RMSE} - Root mean squared error is used to represent the measure of difference between ground truth CO$_2$ saturation image and the posterior mean of the samples generated by the trained network.

\textbf{Normalized std} - It represents normalized point-wise standard deviation or mean-normalized standard deviation. It is calculated by stabilized division of the standard deviation by the envelope of the conditional mean \cite{siahkoohi2023reliable}. It is used to avoid the bias from strong amplitudes in the estimated image.

\end{appendices}

\end{document}